\documentclass[
12pt,
preprint,preprintnumbers,nofootinbib,
groupedaddress,superscriptaddress,amsmath,amssymb]{revtex4}
\usepackage{graphicx}
\usepackage{dcolumn}
\usepackage{bm}
\usepackage{amssymb}
\usepackage{amsmath}
\usepackage{epsfig}    
\usepackage{color}
\usepackage{slashed}
\usepackage{hhline}

\def\be{\begin{equation}}
\def\ee{\end{equation}}
\newcommand{\bea}{\begin{eqnarray}}
\newcommand{\eea}{\end{eqnarray}}
\newcommand{\nn}{\nonumber}

\numberwithin{equation}{section}

\begin{document}

\title{Neutrino mass generation with large $SU(2)_L$ multiplets \\
under local $U(1)_{L_\mu - L_\tau}$ symmetry}
\preprint{KIAS-P18040}
\author{Takaaki Nomura}
\email{nomura@kias.re.kr}
\affiliation{School of Physics, KIAS, Seoul 02455, Korea}

\author{Hiroshi Okada}
\email{macokada3hiroshi@cts.nthu.edu.tw}
\affiliation{Physics Division, National Center for Theoretical Sciences, Hsinchu, Taiwan 300}

\date{\today}

\begin{abstract}
{We propose a model of neutrino mass matrix with large $SU(2)$ multiplets and gauged $U(1)_{L_\mu - L_\tau}$ symmetry,
in which we introduce $SU(2)$ quartet scalar and quintet fermions with nonzero $L_\mu - L_\tau$ charge.
 Then we investigate the neutrino mass structure and explore phenomenologies of large multiplet fields at the collider,
 particularly, focussing on doubly- and singly- charged exotic leptons from the quintet.}
\end{abstract} 
\maketitle
\newpage

\section{Introduction}
The mass spectrum and flavor structure of fermions are mysterious issues in the standard model (SM).
In particular, at least two non-zero neutrino masses require the existence of physics beyond the SM for its generating mechanism.
Actually there are many mechanisms to generate neutrino mass such as Type-I, II and III seesaw models as well as radiatively induced neutrino mass models. 
One interesting scenario is to generate neutrino mass via interaction among the SM lepton and the exotic fields which are large $SU(2)$ multiplets like quartet, quintet or septet~\cite{Kumericki:2012bh, Law:2013gma, Yu:2015pwa, Nomura:2016jnl,Nomura:2017abu, Wang:2016lve}.
Introducing a quartet scalar and quintet fermions, we can realize neutrino mass generation 
by type-III like seesaw way using the quintet fermions {and the quartet scalar 
with vacuum expectation value (VEV).}
Remarkably, tiny neutrino mass is realized by two suppression effects; Majorana mass of the quintet fermion and small VEV of the quartet scalar {required by the constraint from $\rho$-parameter, which is similar to the type-II seesaw model.}

The other hint of new physics is anomalous magnetic dipole moment of muon (muon $g-2$), where the observed value~\cite{Bennett:2006fi, Patrignani:2016xqp} deviates from the SM prediction~\cite{Davier:2010nc, Jegerlehner:2011ti, Hagiwara:2011af, Aoyama:2012wk} as $\Delta a_\mu \equiv \Delta a_\mu^{\mathrm{exp}} - \Delta a_\mu^{\mathrm{th}} = (28.8 \pm 8.0) \times 10^{-10}$ by 3.6 $\sigma$ confidence level.
One of the interesting possibility to resolve the deviation is given by $U(1)_{L_\mu - L_\tau}$ gauge symmetry, where the associated $Z'$ boson contributes to muon $g-2$~\cite{Gninenko:2001hx,Gninenko:2014pea,Altmannshofer:2016brv}.
 In addition to the muon $g-2$ issue, this gauge symmetry has some attractive properties; gauge anomaly is canceled~\cite{He:1990pn,Foot:1994vd}, lepton flavor non-universality in semileptonic $B$-meson decays can be addressed with some extensions~\cite{Altmannshofer:2014cfa,Crivellin:2015mga,Altmannshofer:2016jzy,Ko:2017yrd,Chen:2017usq,Baek:2017sew}, and other interesting studies have been done in refs.~\cite{Heeck:2014qea,Baek:2015fea,Heeck:2016xkh,Altmannshofer:2016oaq,Patra:2016shz,Biswas:2016yan,Heeck:2011wj, Nomura:2018yej,Chen:2017gvf,Asai:2017ryy, Nomura:2018vfz, Baek:2015mna, Lee:2017ekw,Bauer:2018onh,Biswas:2017ait, Kaneta:2016uyt, Araki:2017wyg,Chen:2017cic}.
 Furthermore $U(1)_{L_\mu - L_\tau}$ symmetry could restrict structure of neutrino mass matrix, since leptons have flavor dependent charges under the symmetry~\cite{Baek:2015mna, Asai:2017ryy, Nomura:2018vfz,Lee:2017ekw}.
 
 In this letter, we construct a model of neutrino mass generation with large $SU(2)$ multiplets and gauged $U(1)_{L_\mu - L_\tau}$ symmetry in which we introduce 
 $SU(2)$ quartet scalar and fermion quintets with $L_\mu - L_\tau$ charge.
 Then we investigate neutrino mass structure of the model and explore the phenomenology of large multiplet fields at the collider experiments.
 In particular we focus on doubly- and singly- charged exotic leptons from the quintet.

This paper is organized as follows.
In Sec.~II, we introduce our model, derive some formulas of active neutrino mass matrix, and show the typical order of Yukawa couplings and related masses.
In Sec.~III, we discuss implications to physics at the {Large Hadron Collider(LHC)} focusing on pair production of  charged particles in the multiplets.
We conclude and discuss in Sec.~IV.

\section{ Model setup}
 \begin{widetext}
\begin{center} 
\begin{table}
\begin{tabular}{|c||c|c|c|c|c|c|c|c|c||c|c|c|c|}\hline\hline  
&\multicolumn{9}{c||}{Lepton Fields} & \multicolumn{3}{c|}{Scalar Fields} \\\hline
& ~$L_{L_e}$~ & ~$L_{L_\mu}$~ & ~$L_{L_\tau}$~ & ~$e_R^{}$~& ~$\mu_R^{}$~ & ~$\tau_R^{}$~ & ~$\Sigma_{R_e}$ ~ & ~$\Sigma_{R_\mu}$ ~ & ~$\Sigma_{R_\tau}$ ~ & ~$H$ ~   & ~$\Phi_4$ & ~$\varphi$~ \\\hline 
$SU(2)_L$ & $\bm{2}$  & $\bm{2}$ & $\bm{2}$  & $\bm{1}$ & $\bm{1}$ & $\bm{1}$  & $\bm{5}$ & $\bm{5}$ & $\bm{5}$ & $\bm{2}$ & $\bm{4}$ & $\bm{1}$  \\\hline 
$U(1)_Y$ & $-\frac12$ & $-\frac12$ & $-\frac12$ & $-1$ & $-1$ & $-1$  & $0$ & $0$ & $0$  & $\frac12$ & $\frac12$ & $0$   \\\hline
$U(1)_{L_\mu - L_\tau}$ & $0$ & $1$ & $-1$ & $0$ & $1$ & $-1$ & $0$  & $1$ & $-1$  & $0$ & $0$ & $1$ \\ \hline 
\end{tabular}
\caption{Contents of fermion and scalar fields
and their charge assignments under $SU(2)_L\times U(1)_Y$.}
\label{tab:1}
\end{table}
\end{center}
\end{widetext}

In this section, we introduce our model based on $U(1)_{L_\mu - L_\tau}$ gauge symmetry, and derive a formula of active neutrino mass matrix.
The particle contents with charge assignments are shown in Tab.~\ref{tab:1}. In fermion sector, we introduce {three} right-handed exotic fermions $\Sigma_R$ which are $SU(2)$ quintet with hypercharge $Y=0$. 
\footnote{{$\Sigma_{R_e}$ is needed to reproduce the current neutrino oscillation data, although this field is not needed for gauge anomaly cancellations.}}
In scalar sector, we introduce $SU(2)$ quartet $\Phi_4$ with hypercharge $Y=1/2$ and the SM singlet $\varphi$ with $U(1)_{L_\mu - L_\tau}$ charge 1.
Under the gauge symmetries, we can write following Yukawa interactions associated with the SM leptons and exotic fermions, and scalar potential~\footnote{Except the $U(1)_{L_\mu-L_\tau}$ gauge symmetry, the field contents are same as the one of ref.~\cite{Kumericki:2012bh}.}:
\begin{align}
& -\mathcal{L}_{Y}
= (y_{\ell})_{ii} \bar L_{L_i} H e_{R_i} +(y_{\nu})_{ij} [\bar L_{L_i} \tilde\Phi_4 \Sigma_{R_j} ] 
+  (M_{\Sigma})_{ee} [\bar \Sigma^c_{R_e} \Sigma_{R_e}] +  (M_{\Sigma})_{\mu \tau} [\bar \Sigma^c_{R_\mu} \Sigma_{R_\tau}]\nonumber \\
& \qquad \qquad + Y_{e \mu} \varphi^* [\bar \Sigma^c_{R_e} \Sigma_{R_\mu}] + Y_{e \tau} \varphi [\bar \Sigma^c_{R_e} \Sigma_{R_\tau}]  + {\rm h.c.} \\
& \mathcal{V} = -\mu_H^2 H^\dagger H  -\mu_\varphi^2 \varphi^* \varphi + M_4^2 \Phi_4^\dagger \Phi_4  + \lambda_H (H^\dagger H)^2 + \lambda_\varphi (\varphi^* \varphi)^2  + (\lambda_{H\Phi_4} [H^\dag \Phi_4 H^\dag \Phi_4] + {\rm h.c.}) 
\nonumber \\
& \qquad \qquad + ( \lambda_0 [H^\dag \Phi_4^* H H] + {\rm h.c.}) + \lambda_{H\varphi} (H^\dagger H)(\varphi^* \varphi) + \mathcal{V}_{\rm trivial\ term},
\label{Eq:lag-flavor}
\end{align}
where $\mathcal{V}_{\rm trivial\ term}$ includes trivial quartic terms,
{and $\lambda_0$ plays a role in evading dangerous Goldstone Boson(GB) from $\Phi_4$} {as well as in inducing the VEV of $\Phi_4$ when $M_4^2 >0$.} 

{\it Scalar sector}: The scalar fields can be written as 
\begin{align}
H =\left[
\begin{array}{c}
w^+\\
\frac{v+ \tilde{h}+iz}{\sqrt2}
\end{array}\right],\quad 
\Phi_4 =\left[\phi^{++}, \phi_2^{+}, \frac{v_4 + \phi_R + i \phi_I}{\sqrt{2}},\phi_1^{-}\right]^T, \quad
\varphi = \frac{1}{\sqrt{2}} (v_\varphi + \tilde{\varphi}_R + i z'),
\label{component}
\end{align}
where $w^+$, $z$ and $z'$ are Nambu-{GB} (NGB) absorbed by $W^+$, $Z$ and $Z'$ bosons, and $v$, $v_4$ and $v_\varphi$ are VEVs of each field.
The VEVs are obtained by applying the conditions $\partial \mathcal{V}/\partial v = \partial \mathcal{V}/\partial v_4 = \partial \mathcal{V}/\partial v_\varphi =0$, where we take $\{M_4^2, \mu_H^2, \mu_\varphi^2 \} > 0$ in our scaler potential. Then we have
\begin{equation}
v \simeq \sqrt{\frac{\mu_H^2}{\lambda_H}}, \quad v_\varphi \simeq \sqrt{\frac{\mu_\varphi^2}{\lambda_\varphi}}, \quad v_4 \simeq \frac{\lambda_0 v^3}{\sqrt{6} M_4^2},
\end{equation}
where we assumed  $v_4 << \{v, v_\varphi \}$, $\lambda_{H \varphi} \ll 1$ and couplings in the $V_{\rm trivial \ term}$ are small.
The VEV of $\Phi_4$ is restricted by the $\rho$-parameter which is given by 
\begin{align}
\rho=\frac{v^2+7 v_4^2}{v^2+  v_4^2},
\end{align}
where the experimental value is given by $\rho=1.0004^{+0.0003}_{-0.0004}$ at $2\sigma$ confidence level.  
Therefore we should require $v_4 \lesssim 2.65$ GeV, and this bound is naturally satisfied; $v_4 \sim 1$ GeV with  $M_4 \sim 1$ TeV and $\lambda_0 \sim 0.1$.
Assuming small contribution from terms in $V_{\rm trivial\ term}$, the masses for components in $\Phi_4$ is given by $M_4$. 
The squared mass terms for CP-even scalar bosons $\{ \tilde h, \tilde \varphi_R \}$ are given by
\begin{equation}
\mathcal{L} \supset \frac{1}{4} \begin{pmatrix} \tilde h\\ \tilde \varphi_R \end{pmatrix}^T \begin{pmatrix} \lambda_H v^2 & \lambda_{H \varphi} v v_\varphi \\  \lambda_{H \varphi} v v_\varphi  & \lambda_\varphi v_\varphi^2 \end{pmatrix} \begin{pmatrix} \tilde h \\ \tilde \varphi_R \end{pmatrix},
\end{equation} 
where the mixing with neutral component of $\Phi_4$ is negligibly small due to small $v_4$.
The above squared mass matrix can be diagonalized by an orthogonal matrix and the mass eigenvalues are given by
\begin{equation}
m_{h,\phi}^2 = \frac{\lambda_H v^2 +\lambda_\varphi v_\varphi^2 }{4} \pm \frac{1}{4} \sqrt{\left( \lambda_H v^2 -\lambda_\varphi v_\varphi^2 \right)^2 + 4 \lambda_{H \varphi}^2 v^2 v_\varphi^2 }.
\end{equation}
The corresponding mass eigenstates $h$ and $\varphi_R$ are obtained as   
\begin{equation}
\begin{pmatrix} h \\ \varphi_R \end{pmatrix} = \begin{pmatrix} \cos \alpha & \sin \alpha \\ - \sin \alpha & \cos \alpha \end{pmatrix} \begin{pmatrix} \tilde H \\ \tilde \varphi_R \end{pmatrix}, \quad
\tan 2 \alpha = \frac{2 \lambda_{H \varphi} v v_\varphi}{\lambda_H v^2 - \lambda_\varphi v_\varphi^2},
\label{eq:scalar-mass-fields}
\end{equation}
where $\alpha$ is the mixing angle, and $h$ is identified as the SM-like Higgs boson when $\alpha \ll 1$.
For later convenience, we write gauge interactions giving decay process of $\Phi_4$ components such that 
\begin{align}
|D_\mu \Phi_4|^2 \supset \frac{1}{8} \frac{g_2^2 }{c_W^2} v_4 Z_\mu Z^\mu \phi_R + g_2^2 v_4 W^+_{\mu} W^{- \mu} \phi_R +  \sqrt{\frac{3}{2}} v_4 g_2^2 W^\pm W^\pm \phi^{\mp \mp} + \nonumber \\
+\frac{g_2^2 v_4}{c_W} \left[ s_W^2 Z_\mu W^{+ \mu} \phi^-_2 + \frac{\sqrt{3}}{2}(2 - s_W^2 )    Z_\mu W^{+ \mu } \phi^-_1 + c.c. \right],
\label{eq:Phi4int}
\end{align} 
where $c_{W}(s_W) = \cos \theta_W (\sin \theta_W)$ with the Weinberg angle $\theta_W$ and $g_2$ is the $SU(2)_L$ gauge coupling constant.

{\it Z' boson and muon $g-2$}: 
After $\varphi$ developing VEV, $U(1)_{L_\mu - L_\tau}$ symmetry is spontaneously broken resulting in massive $Z'$ boson.
We obtain $Z'$ boson such as 
\begin{equation}
m_{Z'} = g_X v_\varphi,
\end{equation}
where $g_X$ is the gauge coupling constant associated with $U(1)_{L_\mu - L_\tau}$ and we have ignored $U(1)$ kinetic mixing assuming it is negligibly small.
Gauge interactions among $Z'$ and the SM fermions are given by 
\begin{equation}
g_X Z'_\nu (\bar L_\mu \gamma^\nu L_\mu - \bar L_\tau \gamma^\nu L_\tau + \bar \mu_R \gamma^\nu \mu_R - \bar \tau_R \gamma^\nu \tau_R).
\end{equation}
The $Z'$ contribution to muon $g-2$ is estimated as
\begin{equation}
 \Delta a_\mu^{Z'} = \frac{g_X^2}{8 \pi^2} \int_0^1 dx \frac{2 m_\mu^2 x^2 (1-x)}{x^2 m_\mu^2 + (1-x) m_{Z'}^2}.
\end{equation}
We can explain muon $g-2$ with $m_{Z'} \sim \mathcal{O}(0.1)$ GeV and $10^{-4}  \lesssim g_X \lesssim 10^{-3}$ without conflict to other constraints.
In this region $Z'$ dominantly decays into neutrino pair.

{\it Fermion quintet}:
After $\varphi$ developing a VEV, the Majorana mass matrix for $\Sigma_{R_i}$ is obtained as
\begin{equation}
M_\Sigma = \begin{pmatrix} (M_{\Sigma})_{ee} & \frac{Y_{e\mu} v_\varphi}{\sqrt{2}} & \frac{Y_{e\tau} v_\varphi}{\sqrt{2}} \\
\frac{Y_{e\mu} v_\varphi}{\sqrt{2}} & 0 & (M_\Sigma)_{\mu \tau} \\ \frac{Y_{e\tau} v_\varphi}{\sqrt{2}} & (M_\Sigma)_{\mu \tau} & 0 \end{pmatrix}. \label{eq:Msigma}
\end{equation}
Here we write the quintet Majorana fermions by components such that
\begin{align} 
\Sigma_R&
\equiv
\left[\Sigma_1^{++},\Sigma_1^{+},{\Sigma^0}, \Sigma_{2}^-, \Sigma_{2}^{--} \right]_R^T, 
\label{eq:sigmaR}
\end{align}
where the upper indices represent the electric charges and lower indices distinguish components with the same electric charge for each component.
The quintet is also written as $(\Sigma_{R_a})_{ijkl}$ where the indices $\{i,j,k,l \}$ take $1$ or $2$ corresponding to $SU(2)_L$ doublet index~\footnote{Here $(\Sigma_R)_{ijkl}$ is the symmetric tensor notation which is explicitly given by $(\Sigma_R)_{[1111]} = \Sigma_{1R}^{++}$, $(\Sigma_4)_{[1112]} = \Sigma_{1R}^{+}/\sqrt{3}$, $(\Sigma_R)_{[1122]} = \Sigma^{0}_R/\sqrt{3}$, $(\Sigma_R)_{[1222]} = \Sigma^{-}_{2R}$ and $(\Sigma_R)_{[2222]} = \Sigma^{--}_{2R}$.}.
The mass term can be expanded as 
\begin{align}
M_{\Sigma_{ab}} [\bar \Sigma_{R_a}^c \Sigma_{R_b}] &= M_{\Sigma_{ab}} (\bar \Sigma_{R_a}^c)_{ijkl} ( \Sigma_{R_b})_{i'j'k'l'} \epsilon^{ii'} \epsilon^{jj'} \epsilon^{kk'} \epsilon^{ll'} \nonumber \\
& = M_{\Sigma_{ab}} \left[ \bar \Sigma_{1R_a}^{++ c} \Sigma_{2R_b}^{--} + \bar \Sigma_{1R_a}^{+ c} \Sigma_{2R_b}^{-} + \bar \Sigma_{2R_a}^{- c} \Sigma_{1R_b}^{+} + \bar \Sigma_{2R_a}^{-- c} \Sigma_{1R_b}^{++} + 
\bar \Sigma_{Ra}^{0 c} \Sigma_{R_b}^{0} \right] \nonumber \\
& = M_{\Sigma_{ab}} [\bar \Sigma^{++}_a \Sigma^{++}_b + \bar \Sigma^{+}_a \Sigma^{+}_b + \Sigma^{0c}_{R_a} \Sigma^0_{R_b}],  
\end{align}
where $\epsilon^{ii'} (i,i' =1,2)$ is the antisymmetric tensor acting on $SU(2)$ representation space, and we rewrite components as $\Sigma^{++(+)}_{1R} = \Sigma^{++(+)}_R$ and $\Sigma^{--(-)c}_{2R} = \Sigma^{++(+)}_L$. 
Notice that $\Sigma_{1}^{\pm (\pm \pm)}$ and $\Sigma_{2}^{\pm (\pm \pm)}$ are combined to compose singly(doubly)- charged Dirac fermions while $\Sigma^0_R$ remain as neutral Majorana fermion.
The mass eigenvalues of each component are given by diagonalizing mass matrix $M_{\Sigma_{ab}}$ where mixing between the SM leptons will be negligibly small due to small VEV of $\Phi_4$.
Then mass matrix can be diagonalized via $M_\Sigma^{\rm diagonal} = V M_{\Sigma} V^T$ with mixing matrix $V$.

\subsection{Neutrino mass matrix}

\begin{figure}[tb]\begin{center}
\includegraphics[width=70mm]{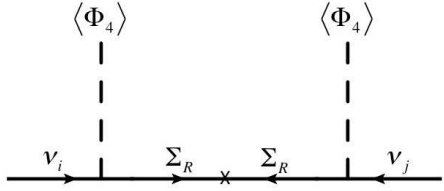}
\caption{The diagram for neutrino mass generation.
}   \label{fig:neut1}\end{center}\end{figure}

Here we derive formula for the active neutrino mass matrix. 
Firstly the relevant interaction in Yukawa coupling is given by 
\begin{align}
-{\cal L}& \supset (y_\nu)_{aa}
\left[
\bar\nu_{L_a} \left(\frac{1}{\sqrt2}\Sigma_{R_a}^0 \phi^{0*} +\frac{\sqrt3}{2} \Sigma_{1R_a}^+ \phi_1^- +  \frac{1}{2}\Sigma_{2R_a}^- \phi_{2}^+ +  \Sigma_{1R_a}^{++} \phi^{--}  \right)\right.
\nn\\
&\left. +
\bar\ell_{L_a} \left(\frac{1}{\sqrt2}\Sigma_{R_a}^0 \phi_1^-  + \frac12 \Sigma_{1R_a}^+ \phi^{--} +  \frac{\sqrt3}{2}\Sigma_{2R_a}^- \phi^{0*} +  \Sigma_{2R_a}^{--} \phi_{2}^{+}  \right)\right] +{\rm h.c.}
\nn\\& \supset 
 \frac{ {(y_\nu)_{aa}}}{\sqrt2}
\bar\nu_{L_a} \Sigma_{R_a}^0 \phi^{0*} +{\rm h.c.},
\label{eq:Yukawa}
\end{align}
where the terms in the last line contribute to generate the active neutrino mass matrix.
Then the active neutrino mass matrix $m_\nu$ is generated as Fig.~\ref{fig:neut1} whose formula is given by
\begin{align}
 (m_{\nu})_{ij}
=\sum_{a,b=1}^3 (y_\nu)_{ia} (M_{\Sigma}^{-1})_{ab} (y_\nu^T)_{bj}v_4^2.
\end{align}
{If we take the scale of neutrino mass is $\mathcal{O}(0.1)$ eV the magnitude of the Yukawa coupling $y_\nu$ is around $10^{-3}$ for $v_4 \sim 1$ GeV and $M_\Sigma \sim 1$ TeV. 
Thus we can obtain tiny neutrino mass naturally as the $y_\nu$ is similar size as the muon Yukawa coupling in the SM.}
In addition, from the (inverse) two-zero texture analysis, the above neutrino mass matrix provides several predictions such as correlation among neutrino mass eigenvalues and CP-phase; for detailed analysis see, e.g., ref.~\cite{Asai:2017ryy}.
\footnote{In general, one-loop induced neutrino mass matrix is also induced via $\lambda_{H\Phi_4}$. But here we simply neglect this term, assuming $\lambda_{H\Phi_4}<<1$. Even when  this contribution is comparable to the tree level,  our prediction retains if $M_\Sigma$ is degenerate to the neutral component of $\Phi_4$~\cite{Ma:2006km}.}

\subsection{Beta function of $g_2$}
\label{beta-func}
\begin{figure}[tb]
\begin{center}
\includegraphics[width=13cm]{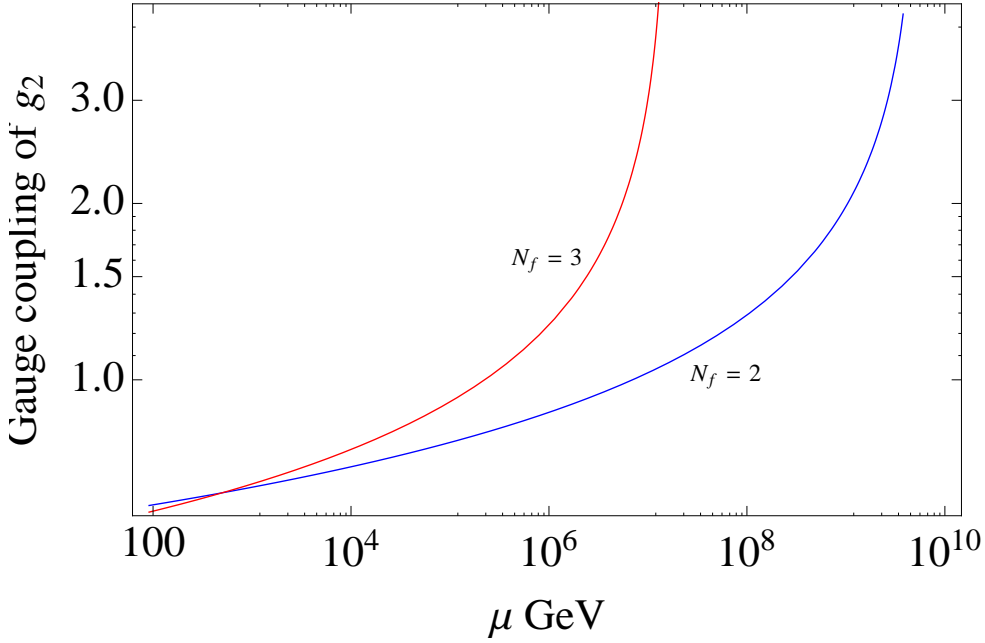}
\caption{The running of $SU(2)_L$ gauge coupling $g_2$ in terms of a reference energy scale $\mu$.}
\label{fig:rge}
\end{center}\end{figure}

Here we discuss running of gauge coupling and estimate the effective energy scale by evaluating the Landau pole for $SU(2)_L$ gauge coupling $g_2$ in the presence of new large $SU(2)_L$ multiplet fields with nonzero hypercharges~\footnote{We have confirmed that the gauge coupling of $U(1)_Y$ is relevant up to Planck scale.}.
The new contributions to $g_2$ from one $SU(2)_L$ quintet fermion ($\Sigma_R$) and  quartet boson ($\Phi_4$) are given by
\begin{align}
\Delta b^{\Sigma_R}_{g_2}=\frac{20}{3}, \quad \Delta b^{\Phi_4}_{g_2}=\frac{5}{3} .
\end{align}
Then one finds that the energy evolution of the gauge coupling $g_2$ as~\cite{Nomura:2017abu, Kanemura:2015bli}
\begin{align}
\frac{1}{g^2_{g_2}(\mu)}&=\frac1{g_2^2(m_{in.})}-\frac{b^{SM}_{g_2}}{(4\pi)^2}\ln\left[\frac{\mu^2}{m_{in.}^2}\right]\nn\\
&
-N_f \theta(\mu-m_{th.}) \frac{\Delta b^{\Sigma_R}_{g_2}}{(4\pi)^2}\ln\left[\frac{\mu^2}{m_{th.}^2}\right]
-\theta(\mu-m_{th.}) \frac{\Delta b^{\Phi_4}_{g_2}}{(4\pi)^2}\ln\left[\frac{\mu^2}{m_{th.}^2}\right],\label{eq:rge_g}
\end{align}
where $N_f$ is the number of $\Sigma_R$, $\mu$ is a reference energy, $b^{SM}_{g_2}=-19/6$, and we assume $m_{in.}(=m_Z)<m_{th.}=$500 GeV, being respectively threshold masses of exotic fermions and bosons for $m_{th.}$.
The resulting flow of ${g_2}(\mu)$ is then given by the Fig.~\ref{fig:rge} for $N_f =2$ and $N_f =3$.
This figure shows that $g_2$ is relevant up to the mass scale $\mu={\cal O}(10^3\sim 10^4)$ TeV in case of $N_f=3$,
while $g_2$ is relevant up to the mass scale $\mu={\cal O}(10^6)$ TeV in case of $N_f=2$.
Thus our theory does not spoil, as far as we work on at around the scale of TeV.

\section{Phenomenology}

In this section, we discuss phenomenology of the model focusing on interactions among quintuplet fermions and $Z'$ boson~\footnote{Collider phenomenology of quartet scalar is discussed in refs.~\cite{delAguila:2013yaa,delAguila:2013mia, Nomura:2017abu,Chala:2018ari}.}.
The relevant interaction is written as 
\begin{align}
\mathcal{L} \supset & g_{X} (V_{i2} (V^T)_{2j} - V_{i3} (V^T)_{3j}) Z'_\mu \left[ \bar \Sigma_i^{++} \gamma^\mu \Sigma_j^{++} + \bar \Sigma_i^{+} \gamma^\mu \Sigma_j^{+} + \bar \Sigma_i^{0} \gamma^\mu P_R \Sigma_j^{0} \right] \nonumber \\
& \equiv X_{ij} Z'_\mu \left[ \bar \Sigma_i^{++} \gamma^\mu \Sigma_j^{++} + \bar \Sigma_i^{+} \gamma^\mu \Sigma_j^{+} + \bar \Sigma_i^{0} \gamma^\mu P_R \Sigma_j^{0} \right]
\end{align}
where $\Sigma_i^{Q}$ denotes a mass eigenstate with electric charge $Q$.
 For illustration, here we show $X_{ij}$ in the case of $(M_{\Sigma})_{11} = 550$ GeV, $(M_{\Sigma})_{12(21)} = 830$ GeV, $(M_{\Sigma})_{13(31)} = 950$ GeV and $(M_{\Sigma})_{23(32)} = 830$ GeV
in Eq.~(\ref{eq:Msigma}), which provides mass eigenvalues of $\Sigma$ as $(622, 833, 2006)$ GeV. In this case, we obtain 
\begin{equation}
X_{ij} = g_X \begin{pmatrix} 0.17 & -0.64 & 0.088 \\ -0.64 & -0.15 & -0.74 \\ 0.088 & -0.74 & -0.012 \end{pmatrix}_{ij}.
\end{equation}
This matrix determines dominant mode in decay of $\Sigma^{Q}_i \to \Sigma^Q_{j} Z'$; for example $\Sigma^{Q}_3 \to \Sigma^{Q}_2 Z'$ mode dominate $\Sigma^{Q}_3 \to \Sigma^{Q}_1 Z'$ mode in this $X_{ij}$.
The components of the quintuplets can be produced via electroweak gauge interaction which is given by
\begin{align}
\mathcal{L} \supset &  -\bar \Sigma^{++}_i \gamma^\mu ( i 2 e A_\mu + i 2 g_2 c_W Z_\mu) \Sigma^{++}_i -\bar \Sigma^{+}_i \gamma^\mu (i e A_\mu + i g_2 c_W Z_\mu) \Sigma^{+}_i \nonumber \\
& +\left( i \sqrt{2}  g_2 \bar \Sigma^{++}_i \gamma^\mu W_\mu^+ \Sigma^+_i + i \sqrt{3} g_2 \bar \Sigma^+_i \gamma^\mu W^+_\mu P_R \Sigma_i^0 + h.c. \right)
\end{align}
where these terms are diagonal for generations of the quintuplets.

\begin{figure}[tb]\begin{center}
\includegraphics[width=70mm]{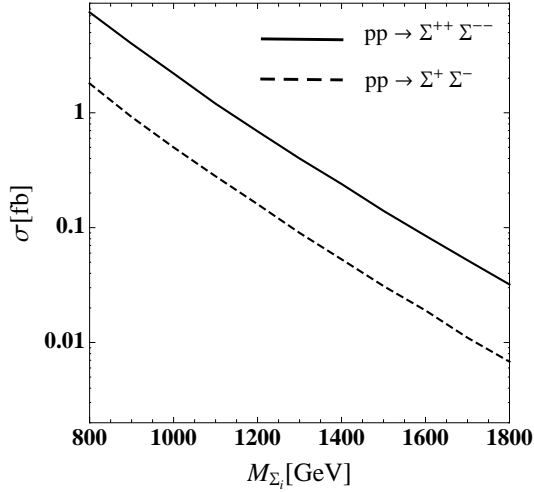}
\caption{The cross section for pair production process $p p \to Z/\gamma \to \Sigma^{++}_i \Sigma^{--}_i(\Sigma^+_i \Sigma^-_i)$ as a function of their mass.
}   \label{fig:CXsigma}\end{center}\end{figure}
Here we estimate the cross section of production process $pp \to \Sigma^{++}_i \Sigma^{--}_i(\Sigma^{+}_i \Sigma^{-}_i)$ via electroweak interaction using {\tt CalcHEP}~\cite{Belyaev:2012qa} 
implementing relevant interactions with {\tt CTEQ6L} PDF~\cite{Nadolsky:2008zw}.
In Fig.~\ref{fig:CXsigma}, we show the cross sections for pair production as a function of exotic fermion mass.
We find that production cross section for doubly-charged fermion can be $\mathcal{O}(1)$ fb for $m_{\Sigma_i} = 1$ TeV while that for singly-charged fermion is smaller.

The components $\Sigma_i^Q$ decay through gauge interactions and Yukawa interactions in Eq.~(\ref{Eq:lag-flavor}).
 Firstly decay mode  $\Sigma_i^Q \to \Sigma_i^{Q \mp 1} W^\pm$ mode with on-shell $W^\pm$ is not kinematically allowed 
since initial and final state exotic fermion masses are degenerated at tree level~\footnote{Mass difference can appear through mixing among SM leptons but it is negligibly small}.
At one-loop level, mass difference is induced which can be few $100$s MeV scale. 
Then decay mode of $\Sigma_i^Q \to \Sigma_i^{Q \mp 1} \pi^\pm$ is kinematically allowed which is induced via off-shell $W^\pm$ boson; $\Sigma_i^{Q \mp 1} \ell^\pm \nu$ mode is subdominant.
The decay width is given by
\begin{equation}
\Gamma_{\Sigma_i^Q \to \Sigma_i^{Q \mp 1} \pi^\pm} = \frac{6 G_F^2 V_{ud}^2 \Delta M^3 f_\pi^2}{\pi} \sqrt{1 - \frac{m_\pi^2}{\Delta M^2}},
\end{equation}
where $m_\pi \simeq 140$ MeV is charged pion mass, $f_\pi = 131$ MeV is pion decay constant and $G_F$ is the Fermi constant.
Adopting $\Delta M = 166$ MeV in quintet case, we obtain $\Gamma_{\Sigma_i^Q \to \Sigma_i^{Q \mp 1} \pi^\pm} \sim 10^{-14}$ GeV.
In addition $\Sigma_i^Q$ can decay into "$Z'$ + lighter generation" and/or "SM lepton + component of $\Phi_4$" depending on the generation of the $\Sigma_i^Q$.
The partial decay widths are then given by
\begin{align}
\Gamma_{\Sigma_i^Q \to Z' \Sigma_j^Q} = & \frac{|X_{ij}|^2}{8 \pi} m_{\Sigma_i} \lambda^{\frac12}(m_{\Sigma_i}, m_{\Sigma_j} ,m_{Z'}) \nonumber \\
& \times  \left(1 - \frac{3 m_{\Sigma_j}}{m_{\Sigma_i}} + \frac{m_{\Sigma_i}^2 - m_{\Sigma_j}^2 -2 m_{Z'}^2}{2m_{Z'}^2} \sqrt{\lambda (m_{\Sigma_i}, m_{\Sigma_j} ,m_{Z'})+\frac{m_{Z'}^2}{4 m_{\Sigma_i}^2}} \right) \\
\Gamma_{\Sigma_i^Q \to \ell^\pm(\nu_\ell) \phi^{Q'}} = & \frac{(V_{i \ell} y_{\ell \ell} C_{\phi \Sigma})^2}{16 \pi} m_{\Sigma_i} \left( 1 + \frac{m_{\phi}^4}{m_{\Sigma_i}^4} -2 \frac{m_{\phi}^2}{m_{\Sigma_i}^2} \right)^{2} \\
\lambda(m_1, m_2, m_3) = & 1 + \frac{m_2^4}{m_1^4} + \frac{m_3^4}{m_1^4} -2  \frac{m_2^2 m_3^2}{m_1^4} -2  \frac{m_2^2}{m_1^2} -2 \frac{m_3^2}{m_1^2},
\end{align}
where charge of $\phi^{Q'}$ is determined by final state lepton in second decay mode, and $C_{\phi \Sigma}$ is numerical factor which appear each component of Eq.~(\ref{eq:Yukawa}).
The value of $\Gamma_{\Sigma_i^Q \to \ell^\pm(\nu_\ell) \phi^{Q'}}$ is $\sim 10^{-7}$ GeV taking $m_{\Sigma_i} = 1000$ GeV, $m_\phi =500$ GeV and $V_{i\ell} y_{\ell \ell} = 10^{-4}$, 
and it is much larger than $\Gamma_{\Sigma_i^Q \to \Sigma_i^{Q \mp 1} \pi^\pm}$.
We can thus neglect decay mode with pion in our analysis.
Furthermore we find that the first mode has enhancement factor of $m_{\Sigma_i}^2/{m_{Z'}^2}$ which is large for light $Z'$ case motivated by explaining muon $g-2$.
Thus heavier generation of the quintuplet dominantly decay into lighter generation with $Z'$ boson. 
We also find that the BRs of first generation of quintuplets are given by
\begin{align}
& BR(\Sigma_1^{++} \to \{\phi^+ \ell^+, \phi^{++} \nu \}) = \{0.5, 0.5 \} \\
& BR(\Sigma_1^+ \to \{\phi^+ \nu, \phi^{++} \ell^-, \phi^0 \ell^+ \}) = \{ 0.65, 0.09, 0.26 \}, 
\end{align}
where $\ell^\pm$ and $\nu$ include all lepton flavor, and $\phi^\pm$ includes $\phi_{1,2}^\pm$ and $\phi^0 = (\phi_R+i \phi_I)/\sqrt{2}$.

Here we consider the signal at the LHC where we assume $m_{\Sigma_1} > m_{\Phi_4}$ so that the first generation of the quintuplet decays into $\Phi_4$ component with the SM lepton.
The components of $\Phi_4$ decay into the SM gauge bosons via interaction in Eq.~(\ref{eq:Phi4int}).
Then signal processes of our interest are 
\begin{align}
pp & \to \Sigma^{++}_1 \Sigma^{--}_1, \ ( \Sigma^{++}_1 \to \phi^+ \ell^+ \to Z W^+ \ell^+), \\
pp & \to \Sigma^{+}_1 \Sigma^{-}_1, \ ( \Sigma^{+}_1 \to  \phi^0 \ell^+ \to Z Z \ell^+), \\
pp & \to \Sigma^{++}_2 \Sigma^{--}_2, \ (\Sigma^{++}_2 \to Z' \Sigma^{++}_1 \to Z' \phi^+ \ell^+ \to Z' Z W^+ \ell^+), \\
pp & \to \Sigma^{+}_2 \Sigma^{-}_2, \ (\Sigma^{+}_2 \to Z' \Sigma^{+}_1 \to Z' \phi^0 \ell^+ \to Z' Z Z \ell^+),
\end{align}
where $Z'$ dominantly decays into the SM neutrinos. 
These processes give signals of multi-leptons with/without jets and missing transverse momentum.
In Table.~\ref{tab:2}, we summarize BRs for each final state from decay of $\Sigma_1^{++(+)}$ where final states from $\Sigma_2^{++(+)}$ are given by adding $Z'$ since $BR(\Sigma_2^{++(+)} \to Z' \Sigma_1^{++(+)}) \simeq 1$ as discussed above. 
BRs for multilepton final states are relatively small and we need large integrated luminosity to explore the signal.
Thus these signal could be tested at the High-Luminosity LHC. 
Notice that the masses of doubly- and singly- charged exotic fermion are same in each generation and reconstruction of the mass spectrum is important to confirm our scenario.
However, since each signal has many particles in final state, detailed analysis is beyond the scope of this letter. 
 \begin{widetext}
\begin{center} 
\begin{table}
\begin{tabular}{|c||c|c|c|c||c|c|c|}\hline\hline  
& \multicolumn{4}{c||}{$\Sigma_1^{++}$} & \multicolumn{3}{c|}{$\Sigma_1^+$} \\\hline
final state & $\ell^+ \ell'^+ \ell''^+ \ell'^- \nu$ & $\ell^+ \ell'^+ \ell'^+ jj$ & $\ell^+ \ell''^+ j j \nu$ & $\ell^+ jjjj$ & $\ell^+ \ell'^+ \ell''^+ \ell'^- \ell''^-$ & $\ell^+ \ell'^+ \ell'^- jj$ & $\ell^+ jjjj$ \\
 BR & $0.0070$ & $0.025$ & $0.070$ & $0.25$ & 0.0013 & 0.013 & 0.13 \\ \hline
\end{tabular}
\caption{BRs from decay of $\Sigma_1^{++}$ and $\Sigma_1^+$ where $\ell$ is electron or muon.}
\label{tab:2}
\end{table}
\end{center}
\end{widetext}

\section{Conclusions and discussions}

We have constructed a model of neutrino mass generation via interactions among large multiplet of $SU(2)_L$ under $U(1)_{L_\mu - L_\tau}$ gauge symmetry.
The neutrino mass is given by the Yukawa interaction with the quartet scalar and the quintet fermion.
Then we realize tiny neutrino mass by suppression factors regarding the small quartet VEV and Majorana mass of the quintet.
We find neutrino mass matrix has predictive structure due to the $U(1)_{L_\mu - L_\tau}$ symmetry.

Then collider physics of the model is investigated, where we have focused on doubly- and singly-charged exotic leptons.
We find that heavier generations of them dominantly decay into lighter one and $Z'$ boson, while the lightest one decays into SM lepton and components in quartet scalar.
The components of quartet scalar decay into SM gauge bosons when we assume they are lighter than the quintet fermions.
In such a case, the signal of exotic lepton productions is multi-leptons with/without jets and missing transverse momentum.


\section*{Acknowledgments}
\vspace{0.5cm}
H. O. is sincerely grateful for all the KIAS members.


\begin{thebibliography}{99}


\bibitem{Kumericki:2012bh} 
  K.~Kumericki, I.~Picek and B.~Radovcic,
  Phys.\ Rev.\ D {\bf 86}, 013006 (2012)
  [arXiv:1204.6599 [hep-ph]].
  
\bibitem{Law:2013gma} 
  S.~S.~C.~Law and K.~L.~McDonald,
  Phys.\ Rev.\ D {\bf 87}, no. 11, 113003 (2013)
  [arXiv:1303.4887 [hep-ph]].
  
\bibitem{Yu:2015pwa} 
  Y.~Yu, C.~X.~Yue and S.~Yang,
  Phys.\ Rev.\ D {\bf 91}, no. 9, 093003 (2015)
  [arXiv:1502.02801 [hep-ph]].

\bibitem{Nomura:2016jnl} 
  T.~Nomura, H.~Okada and Y.~Orikasa,
  Phys.\ Rev.\ D {\bf 94}, no. 5, 055012 (2016)
  [arXiv:1605.02601 [hep-ph]].

\bibitem{Wang:2016lve} 
  W.~Wang and Z.~L.~Han,
  JHEP {\bf 1704}, 166 (2017)
  [arXiv:1611.03240 [hep-ph]].

\bibitem{Nomura:2017abu} 
  T.~Nomura and H.~Okada,
  Phys.\ Rev.\ D {\bf 96}, no. 9, 095017 (2017)
  [arXiv:1708.03204 [hep-ph]].
  

  
  
\bibitem{Bennett:2006fi} 
  G.~W.~Bennett {\it et al.} [Muon g-2 Collaboration],
  Phys.\ Rev.\ D {\bf 73}, 072003 (2006)
  [hep-ex/0602035].

\bibitem{Patrignani:2016xqp} 
  C.~Patrignani {\it et al.} [Particle Data Group],
  Chin.\ Phys.\ C {\bf 40}, no. 10, 100001 (2016).


\bibitem{Davier:2010nc} 
  M.~Davier, A.~Hoecker, B.~Malaescu and Z.~Zhang,
  Eur.\ Phys.\ J.\ C {\bf 71}, 1515 (2011)
  Erratum: [Eur.\ Phys.\ J.\ C {\bf 72}, 1874 (2012)]
  [arXiv:1010.4180 [hep-ph]].

\bibitem{Jegerlehner:2011ti} 
  F.~Jegerlehner and R.~Szafron,
  Eur.\ Phys.\ J.\ C {\bf 71}, 1632 (2011)
  [arXiv:1101.2872 [hep-ph]].

\bibitem{Hagiwara:2011af} 
  K.~Hagiwara, R.~Liao, A.~D.~Martin, D.~Nomura and T.~Teubner,
  J.\ Phys.\ G {\bf 38}, 085003 (2011)
  [arXiv:1105.3149 [hep-ph]].
  
\bibitem{Aoyama:2012wk} 
  T.~Aoyama, M.~Hayakawa, T.~Kinoshita and M.~Nio,
  Phys.\ Rev.\ Lett.\  {\bf 109}, 111808 (2012)
  [arXiv:1205.5370 [hep-ph]].

  
  
  
  
  
  
  

\bibitem{He:1990pn} 
  X.~G.~He, G.~C.~Joshi, H.~Lew and R.~R.~Volkas,
  Phys.\ Rev.\ D {\bf 43}, 22 (1991).
   
\bibitem{Foot:1994vd} 
  R.~Foot, X.~G.~He, H.~Lew and R.~R.~Volkas,
  Phys.\ Rev.\ D {\bf 50}, 4571 (1994)
  [hep-ph/9401250].
  
\bibitem{Gninenko:2001hx} 
  S.~N.~Gninenko and N.~V.~Krasnikov,
  Phys.\ Lett.\ B {\bf 513}, 119 (2001)
  [hep-ph/0102222].
  
\bibitem{Gninenko:2014pea} 
  S.~N.~Gninenko, N.~V.~Krasnikov and V.~A.~Matveev,
  Phys.\ Rev.\ D {\bf 91}, 095015 (2015)
  [arXiv:1412.1400 [hep-ph]].
  
\bibitem{Altmannshofer:2016brv} 
  W.~Altmannshofer, C.~Y.~Chen, P.~S.~Bhupal Dev and A.~Soni,
  Phys.\ Lett.\ B {\bf 762}, 389 (2016)
  [arXiv:1607.06832 [hep-ph]].

  


\bibitem{Altmannshofer:2014cfa} 
  W.~Altmannshofer, S.~Gori, M.~Pospelov and I.~Yavin,
  Phys.\ Rev.\ D {\bf 89}, 095033 (2014)
  [arXiv:1403.1269 [hep-ph]].

\bibitem{Crivellin:2015mga} 
  A.~Crivellin, G.~D'Ambrosio and J.~Heeck,
  Phys.\ Rev.\ Lett.\  {\bf 114}, 151801 (2015)
  [arXiv:1501.00993 [hep-ph]].
  


\bibitem{Altmannshofer:2016jzy} 
  W.~Altmannshofer, S.~Gori, S.~Profumo and F.~S.~Queiroz,
  JHEP {\bf 1612}, 106 (2016)
  [arXiv:1609.04026 [hep-ph]].
 
\bibitem{Ko:2017yrd} 
  P.~Ko, T.~Nomura and H.~Okada,
  Phys.\ Rev.\ D {\bf 95}, no. 11, 111701 (2017)
  [arXiv:1702.02699 [hep-ph]].
  
\bibitem{Chen:2017usq} 
  C.~H.~Chen and T.~Nomura,
  Phys.\ Lett.\ B {\bf 777}, 420 (2018)
  [arXiv:1707.03249 [hep-ph]].
  
\bibitem{Baek:2017sew} 
  S.~Baek,
  Phys.\ Lett.\ B {\bf 781}, 376 (2018)
  [arXiv:1707.04573 [hep-ph]].
  
  
 
\bibitem{Kaneta:2016uyt} 
  Y.~Kaneta and T.~Shimomura,
  PTEP {\bf 2017}, no. 5, 053B04 (2017)
  [arXiv:1701.00156 [hep-ph]].

\bibitem{Araki:2017wyg} 
  T.~Araki, S.~Hoshino, T.~Ota, J.~Sato and T.~Shimomura,
  Phys.\ Rev.\ D {\bf 95}, no. 5, 055006 (2017)
  [arXiv:1702.01497 [hep-ph]].

  
\bibitem{Chen:2017cic} 
  C.~H.~Chen and T.~Nomura,
  Phys.\ Rev.\ D {\bf 96}, no. 9, 095023 (2017)
  [arXiv:1704.04407 [hep-ph]].
  
\bibitem{Heeck:2011wj} 
  J.~Heeck and W.~Rodejohann,
  Phys.\ Rev.\ D {\bf 84}, 075007 (2011)
  [arXiv:1107.5238 [hep-ph]].

\bibitem{Heeck:2014qea} 
  J.~Heeck, M.~Holthausen, W.~Rodejohann and Y.~Shimizu,
  Nucl.\ Phys.\ B {\bf 896}, 281 (2015)
  [arXiv:1412.3671 [hep-ph]].
  
\bibitem{Baek:2015fea} 
  S.~Baek,
  Phys.\ Lett.\ B {\bf 756}, 1 (2016)
  [arXiv:1510.02168 [hep-ph]].
  
\bibitem{Heeck:2016xkh} 
  J.~Heeck,
  Phys.\ Lett.\ B {\bf 758}, 101 (2016)
  [arXiv:1602.03810 [hep-ph]].
  
\bibitem{Altmannshofer:2016oaq} 
  W.~Altmannshofer, M.~Carena and A.~Crivellin,
  Phys.\ Rev.\ D {\bf 94}, no. 9, 095026 (2016)
  [arXiv:1604.08221 [hep-ph]].

\bibitem{Patra:2016shz} 
  S.~Patra, S.~Rao, N.~Sahoo and N.~Sahu,
  Nucl.\ Phys.\ B {\bf 917}, 317 (2017)
  [arXiv:1607.04046 [hep-ph]].
  
\bibitem{Biswas:2016yan} 
  A.~Biswas, S.~Choubey and S.~Khan,
  JHEP {\bf 1609}, 147 (2016)
  [arXiv:1608.04194 [hep-ph]].
  
\bibitem{Asai:2017ryy} 
  K.~Asai, K.~Hamaguchi and N.~Nagata,
  Eur.\ Phys.\ J.\ C {\bf 77}, no. 11, 763 (2017)
  [arXiv:1705.00419 [hep-ph]].
  

  

  
\bibitem{Chen:2017gvf} 
  C.~H.~Chen and T.~Nomura,
  arXiv:1705.10620 [hep-ph].
  
\bibitem{Biswas:2017ait} 
  A.~Biswas, S.~Choubey, L.~Covi and S.~Khan,
  JCAP {\bf 1802}, no. 02, 002 (2018)
  [arXiv:1711.00553 [hep-ph]].

\bibitem{Nomura:2018yej} 
  T.~Nomura and T.~Shimomura,
  arXiv:1803.00842 [hep-ph].

\bibitem{Nomura:2018vfz} 
  T.~Nomura and H.~Okada,
  arXiv:1803.04795 [hep-ph].
  
\bibitem{Bauer:2018onh} 
  M.~Bauer, P.~Foldenauer and J.~Jaeckel,
  arXiv:1803.05466 [hep-ph].
  
\bibitem{Lee:2017ekw} 
  S.~Lee, T.~Nomura and H.~Okada,
  Nucl.\ Phys.\ B {\bf 931}, 179 (2018)
  [arXiv:1702.03733 [hep-ph]].
  
  
\bibitem{Baek:2015mna} 
  S.~Baek, H.~Okada and K.~Yagyu,
  JHEP {\bf 1504}, 049 (2015)
  [arXiv:1501.01530 [hep-ph]].


\bibitem{Kanemura:2015bli} 
  S.~Kanemura, K.~Nishiwaki, H.~Okada, Y.~Orikasa, S.~C.~Park and R.~Watanabe,
  PTEP {\bf 2016}, no. 12, 123B04 (2016)
  [arXiv:1512.09048 [hep-ph]].

\bibitem{Ma:2006km} 
  E.~Ma,
  Phys.\ Rev.\ D {\bf 73}, 077301 (2006)
  [hep-ph/0601225].
  
\bibitem{delAguila:2013yaa} 
  F.~del Aguila, M.~Chala, A.~Santamaria and J.~Wudka,
  Phys.\ Lett.\ B {\bf 725}, 310 (2013)
  [arXiv:1305.3904 [hep-ph]].
  
\bibitem{delAguila:2013mia} 
  F.~del \'Aguila and M.~Chala,
  JHEP {\bf 1403}, 027 (2014)
  [arXiv:1311.1510 [hep-ph]].


\bibitem{Chala:2018ari} 
  M.~Chala, C.~Krause and G.~Nardini,
  arXiv:1802.02168 [hep-ph].






\bibitem{Belyaev:2012qa} 
  A.~Belyaev, N.~D.~Christensen and A.~Pukhov,
  Comput.\ Phys.\ Commun.\  {\bf 184}, 1729 (2013)
  [arXiv:1207.6082 [hep-ph]].
  
\bibitem{Nadolsky:2008zw} 
  P.~M.~Nadolsky, H.~L.~Lai, Q.~H.~Cao, J.~Huston, J.~Pumplin, D.~Stump, W.~K.~Tung and C.-P.~Yuan,
  Phys.\ Rev.\ D {\bf 78}, 013004 (2008)
  [arXiv:0802.0007 [hep-ph]].

  
\end{thebibliography}
\end{document}